# Social Factors in P2P Energy Trading Using Hedonic Games

Dan Mitrea, Viorica Chifu, Tudor Cioara, Ionut Anghel, Cristina Pop

Computer Science Department, Technical University of Cluj-Napoca, Memorandumului 28, 400114 Cluj-Napoca, Romania

{dan.mitrea; viorica.chifu; tudor.cioara; ionut.anghel; cristina.pop}@cs.utcluj.ro

**Abstract:** Lately, the energy communities have gained a lot of attention as they have the potential to significantly contribute to the resilience and flexibility of the energy system, facilitating widespread integration of intermittent renewable energy sources. Within these communities the prosumers can engage in peer-to-peer trading, fostering local collaborations and increasing awareness about energy usage and flexible consumption. However, even under these favorable conditions, prosumer engagement levels remain low, requiring trading mechanisms that are aligned with their social values and expectations. In this paper, we introduce an innovative hedonic game coordination and cooperation model for P2P energy trading among prosumers which considers the social relationships within an energy community to create energy coalitions and facilitate energy transactions among them. We defined a heuristic that optimizes the prosumers' coalitions, considering their social and energy price preferences and balancing the energy demand and supply within the community. We integrated the proposed hedonic game model into a state-of-the-art blockchain-based P2P energy flexibility market and evaluated its performance within an energy community of prosumers. The evaluation results on a blockchain-based P2P energy flexibility market show the effectiveness in considering social factors when creating coalitions, increasing the total amount of energy transacted in a market session by 5% compared with other game theory-based solutions. Finally, it shows the importance of the social dimensions of P2P energy transactions, the positive social dynamics in the energy community increasing the amount of energy transacted by more than 10% while contributing to a more balanced energy demand and supply within the community.

## 1. Introduction

The energy transition towards renewables is changing the way we produce, consume, and share energy, setting the stage for decentralized local energy systems [1]. In Europe, the war in Ukraine has accelerated this process because it highlighted the security risks associated with centralized energy production and the energy dependence impact on energy prices, encouraging the governments to take more proactive steps towards decentralized and resilient energy systems [2]. Within such a framework, promising new concepts such as prosumers and energy communities have gained a lot of attention as they have the potential to significantly contribute to the resilience and flexibility of the energy system, facilitating the large-scale rollout of intermittent renewable energy technologies without requiring expensive infrastructure upgrades [3].

A prosumer is an individual household that consumes and produces renewable electricity and may inject the surplus into the grid and withdraw electricity when the self-production is not sufficient [4]. Prosumers may participate in the management of energy communities, by trading energy and providing services like demand flexibility and decentralized energy storage [5]. In this context, peer-to-peer (P2P) energy trading enables direct transactions between prosumers, fostering a sense of community by allowing neighbors to share their energy resources. It empowers prosumers to put a value on their energy flexibility increasing their awareness while at the same time promoting energy self-sufficiency and encouraging local collaboration [6]. P2P energy trading platforms are often implemented using blockchain to record energy transactions transparently and in a tampered-proof manner reinforcing the trust among prosumers. The blockchain's decentralized nature and cryptographic algorithms can provide the needed security for data sharing providing support for the interactions within the energy communities [7]. With the help of IoT devices, real-time data on prosumers' energy production and consumption can be collected, allowing for identifying energy-saving opportunities and optimizing the usage of energy resources using Artificial Intelligence (AI) [8]. Innovative incentivization and financing models enable prosumers to directly trade excess energy with each other, pushing the development of a more economically viable and community-driven decentralized energy system [9].

However, even under these favorable conditions, there is an issue that still needs to be addressed such as the level of engagement of prosumers in P2P energy trading which may have the potential to jeopardize the autonomy and reliability of community energy delivery [10]. Prosumer engagement is seen as one of the effective tools to unlock the potential and P2P energy trading and energy communities, by providing mechanisms that are aligned with their social values and expectations (e.g., relations, prices, and other values). Consideration of social relations may foster a sense of community and cooperation, encouraging prosumers to actively trade their energy inside the energy community [11]. They are more willing to participate in P2P energy trading if they have positive social relationships with other peers within the energy community [12]. They tend to be more open to creating local coalitions with prosumers who have strong social connections, such as friends or family members. This can be particularly important in small communities, where players may already have established social ties [13]. Similarly, prosumers are more likely to find mutually acceptable solutions if they value their positive relationships with each other, thus avoiding the energy transactions among peers with negative relations can also be beneficial. Prosumers may also prefer forming coalitions with other prosumers with whom they have a history of cooperation, as this can help to mitigate the risk of conflicts [14]. Social relations can provide information, and support to newcomers, making P2P trading more accessible, the prosumers being more willing to join in trading with community members who share similar environmental values [11].

At the same time from a technical perspective, a significant body of literature is dedicated to employing cooperative game theory for managing the prosumer participation in P2P energy trading [15]. They provide optimized trading strategies by considering the collective interests of prosumers increasing the engagement levels due to consideration of collective goals [16]. The cooperative games enable prosumers to form coalitions over P2P trading platforms, the reward being provided by the collective actions of the members and not solely by individual actions [17]. Moreover, they ensure a fair distribution of energy resources, ensuring that the benefits derived from P2P energy trading [18]. However, most of the approaches discussed in existing literature tend to overlook social aspects such as the social connections among prosumers within the community in the cooperative game models developed for P2P energy trading.

In this paper, we address the gaps identified in the literature by providing a prosumer coordination and cooperation model over a P2P energy flexibility market set up within an energy community. The cooperation model is based on hedonic games addressing the creation of coalitions of prosumers that can have preferences over the peers they are willing to collaborate with or not or for the coalition they may belong to. The prosumers' preferences are expressed as social connections with other peers within the community, while the created coalitions are optimized, matched for trading, and energy balanced using a genetic heuristic. The model implementation was done in the context of a P2P energy flexibility market based on blockchain and smart contracts. The paper contributes to a broader understanding of P2P energy trading and its social dimensions by showing how social connections can be used to guide the process of coalition formation in P2P energy trading and that is possible to create coalitions that satisfy the social preferences of prosumers, while also minimizing the overall difference between surplus and deficit of energy at the community level.

The novel contributions of the paper are the following:

- A cooperation model for P2P energy trading using a hedonic game that considers the social relationships among prosumers within an energy community to create energy coalitions and facilitate energy transactions among them.
- A heuristic to optimize the coalitions of prosumers in the hedonic game model for P2P energy trading, considering the prosumers' social and price preferences and balancing the energy demand and supply within the community.
- The hedonic game model integration with a state-of-the-art blockchain-based P2P energy flexibility market and evaluation in the context of an energy community of prosumers.

The rest of the paper is structured as follows: Section 2 presents the state of the art on cooperative games for peer-to-peer energy trading considering social connections among prosumers, section 3 presents the hedonic game-based cooperative trading model emphasizing the incorporation of social preferences of prosumers in creation and optimization of coalitions, section 4 presents the relevant results in the context of blockchain-based P2P energy flexibility market, section 5 discusses the impact of various parameters on the hedonic game outcomes while section 6 presents conclusions and future work.

## 2. Related Work

The state-of-the-art game theory applications in peer-to-peer energy trading focus on cooperative games to facilitate market-level interactions and prosumer engagement [15, 19]. They investigate how and why certain prosumers might relax some of their goals, and form market-level coalitions collaborating to achieve a better collective outcome [20]. Cooperative games over peer-to-peer energy trading systems are used to guarantee the efficiency, stability, and safety of community-level energy operations, and enforce operational constraints [17, 21]. Prosumers' primary objective is to reduce their energy expenses and increase their profit while using different update strategies to reduce communication burdens between them [22]. However, a limited number of strategies consider social factors as motivators for encouraging collaboration among prosumers within energy communities and none to our knowledge apply hedonic games to capture the social preferences of prosumers [11-13].

Wang et al. [23] propose a two-level hierarchical incentive mechanism to motivate prosumers to join electricity peak-shifting in a P2P decentralized energy market. The Incentivization is based on prosumers having to meet the energy-shifting values they have agreed upon and a reward penalty approach. Luo et al. [24] use a game-theory-based decentralized trading scheme to separate the original coordination problem, which would be the job of a market coordinator, into several sub-problems for each prosumer. The updates on global prices and quantities are done in a sequential manner, such that the next prosumers can use the latest information to optimize their actions, increasing their economic benefits and reducing electricity costs. Intermittent renewable production might cause instability in P2P energy trading and cooperative games can be used to ensure stability [25]. Opportunistic usage of prosumer batteries for peer-to-peer trading is studied in [20] considering that prosumers want to maximize the usage of renewable. Prosumers can either sell their energy surplus without discharging their battery or use their batteries for trading purposes to maximize their utility by considering a charging / discharging action. To form coalitions, a prosumer first must meet its demand from its solar panels, then calculate its surplus or deficit, and then, based on price thresholds of coalitions and available energy prices will be engaged in P2P trading. Coalitional game theory is used in [26] to find the winning coalitions that will play the game of optimizing the energy transfer within a community of prosumers that aim to operate independently from the main grid. Frequent changes in energy demand and generation are considered as well as a utility function that computes the coalition energy availability when all prosumers act together. The management of energy communities based on variations of demand and supply is addressed in [27]. To join coalitions, each prosumer generates a list of preferences that might change over time and uses the Bayesian theorem to update these beliefs about others. Zu et al. [28] integrate energy trading and energy management, such that prosumers can manage their consumption and schedule their green energy storage. The energy control is done using the Lyapunov Theory, allowing prosumers to independently determine their energy order for each time slot, based on their current energy supply conditions. These allow prosumers to join a coordinated mechanism of influencing their energy consumption and green energy charge/discharge and find the best solution for the entire community. Li et al. [29] use game theory to construct two computationally efficient mechanisms for generating stable coalitions of prosumers in P2P energy trading, one that involves a benefit distribution scheme, and another that deals with a novel pricing mechanism. How prosumers decide which coalition to join is based on a dissatisfaction level, which can change as they join different coalitions, determine potential benefits from other coalitions, or consider the option of remaining independent. In [30] the authors proposed a P2P energy trading scheme to establish a grand energy coalition of prosumers and a suitable incentivization mechanism. Cooperative optimization of energy storage units is required, with the cost function quantizing the energy cost saving. The Shapley Value method is used to define a unique distribution of the total monetary benefit of the grand coalition, to all its prosumers, showing a decrease in the energy price. A model based on game theoretic approaches that incentivize prosumers to actively interact with the smart grid, all while preserving the privacy of participants, is presented in [31]. The game assumes that prosumers first choose a price, then consumers observe the prices and decide the amount of power to be purchased. Quadratic functions are used to model the Nash equilibrium assuming that users are willing to consume as much power as needed to balance the generation. Jin et al. [32] use game theory for P2P energy flexibility trading in energy communities. The leader of the game is the producer, and the followers are the consumers. The leader will propose the trading price, and then, the trading quantity, based on multiple factors such as load profile and demand. The price is continuously adjusted as in a Stackelberg game, to reach an optimal value, by comparing the demand with the supply. The solution is beneficial to the entire community because if the seller sets a high price, the buyer will react by reducing his allocated traded quantity. This is done multiple times until the prices and quantities converge. Lee et al. [33] combine three types of games that are played sequentially until the convergence point is met. An evolutionary game is used among buyers with strategies that might evolve, a non-cooperative game between sellers, and a Stackelberg game between sellers and buyers. In the evolutionary game, the community manager updates all members with a loss function that helps them improve over time.

Behavioral models and social motivational models for prosumers to join P2P energy trading schemes are considered in [34]. The norm activation theory is used to model the peers' behavior and to define multiple stages of behavioral change starting with awareness, then responsibility, and finally, personal norms (i.e., the usage of renewable energy). Psychological factors for prosumers joining energy trading schemes related to trust, an ecosystem-friendly and fair distribution models. Game theory is used to consider them at the community level to minimize the gap in supply and demand. Tushar et al. [35] discuss motivational models and social factors for engaging prosumers in energy trading. Aspects such as attitude, rational-economic, information, or positive reinforcement are considered concerning environmental, social, or economic advantages. A trading scheme is designed based on a canonical coalition game to obtain the best energy price and to target the social aspects. In [36], an incentivization solution is proposed using a two-level game-theoretic approach, where the lower level uses a hedonic coalition formation game for the servers to share their resources. Cluster heads are randomly assigned, and they offer reward pools for workers to join them, with these reward pools being different based on the cluster heads' available budgets. As more workers join a cluster head, the reward gets smaller, so the workers will distribute themselves more intelligently such that their reward is the highest possible. Yap et al. [37] use a motivational game theory-based approach to solving efficient P2P energy trading among prosumers, both for multi-cities and intra-city scenarios. The price of the market is set as the average price for residential, commercial, and industrial customers. A cooperative energy market model using a generalized Nash bargaining scheme is proposed in [38] considering social welfare maximization and optimal energy trading. The network operator can trade with prosumers and the prosumers can also P2P trade among them. The socioeconomic optimization problems are transformed from nonconvex problems to linear ones, using a grid propagation algorithm to increase social welfare and fairness in profit allocation. The challenge in P2P energy trading is designing pricing schemes that motivate prosumers to cooperate and participate in managing network congestion [39]. Long et al. address it by proposing a P2P energy trading solution using cooperative games [40]. The parameters for the game are the prosumer load, the energy quantities bought or sold, energy prices, the prosumer electricity bill, and battery energy. Coalitions are formed driven by the income they receive when trading with the supplier while the Shapley Value method is used to allocate resources and fairly distribute cost savings among prosumers. Malik et al. [25] consider multiple time intervals in energy trading, and various priorities such as energy quantity, geographic location, pricing mechanism, etc. After pairs are created, a grand coalition is generated to maximize social welfare and energy savings. Annual profit and energy reliability index are considered to compute a multi-objective optimization function in [41] and perform planning for P2P and P2G energy trading. Game theory and a particle swarm optimization algorithm are used to form coalitions, to find the optimum sizes of the players, and payoff value showing that the profits are maximized when considering both criteria.

## 3. Hedonic Game for Cooperative Trading

In this section, we present how hedonic games can be used for prosumer coordination over a peer-to-peer energy flexibility market implemented at the energy community level. We describe the cooperation model considering the prosumers' social preferences and how the coalitions are optimized and matched using a genetic heuristic.

### 3.1. Cooperation model with social preferences

To model the problem of creating coalitions in a P2P energy trading session, we use a hedonic game approach in which the prosumers are seen as players, that has a specified set of preferences over all possible coalitions:

$$\rho_i \succcurlyeq \rho_j \; such \; that \; v(\rho_i) > v(\rho_i) \tag{1}$$

The preferences include the social relations between two prosumers (i.e., friendship, enemy, neutral established based on social factors or the degree of trust among prosumers) and preferences related to the energy price within the market context. The set of preferences enables the creation of coalitions that not only satisfy the energy requirements of players but also the social relationships between players in the game. Environmental values or social connections may impact prosumers' decisions in forming coalitions for energy trading within a peer-to-peer market. At the same time, the coalition formation process is guided by the need to minimize the difference between total energy surplus and deficit in the energy community.

We define the hedonic game model for peer-to-peer energy trading using the notions of friends, neutral players, and enemies as preferences of prosumers for cooperation. Any pair of prosumers have established a relationship between them that can be one of the friendships ($F$), neutrality $N$, or enemy ($E$):

$$R(p_i, p_j) = val, i \neq j, i, j \in \{1, n\}, R = \{F, N, E\} \quad (2)$$

$$F \succcurlyeq N \succcurlyeq E \quad (3)$$

A friendship relation between two prosumers $p_i$ and $p_j$, should meet one of the following conditions: (i) they have been engaged in positive interactions in previous P2P transactions, (ii) they share common interests, (iii) they have social connections, such as they are friends or family members, or a strong friendship relation not influenced by minor conflicts or external factors. In our model the friendship relation is the following value $v$:

$$v(F(p_i, p_j)) = 1, i \neq j, i, j \in [1, n] \quad (4)$$

An enemy relationship between two prosumers $p_i$ and $p_j$ should meet one of the following: (i) they have been engaged in negative transactional interactions, such as conflict over goals, or one of the two failed to deliver the flexibility as promised in energy transactions and (ii) they have a negative persistence that is unlikely to change due to minor positive interactions or external factors. In our model, the enemy relationship has the following value $v$:

$$v(E(p_i, p_j)) = -1, i \neq j, i, j \in [1, n] \quad (5)$$

A neutral relationship is established between two prosumers $p_i$ and $p_j$ only if all of the following conditions are satisfied: (i) they have not been engaged in a sufficient number of positive interactions in the game to establish a friendship relation or lack strong negative interaction and (ii) they do not express any particular interest or preference towards each other in the game, and their interactions are primarily focused on achieving their own goals. The neutrality relationship is modeled as:

$$v(N(p_i, p_j)) = 0, i \neq j, i, j \in [1, n] \quad (6)$$

The nature of the relationship among prosumers within a coalition is utilized to calculate or assign the hedonic value $V_{rel}$ for the coalition, $C_k$:

$$V_{rel}(C_k) = \sum_{i=1}^{n} \sum_{j=1}^{n} [F(p_i, p_j) + E(p_i, p_j) + N(p_i, p_j)], \forall p_i, p_j \in C_k \quad (7)$$

The preferences related to energy prices are different based on the prosumer role in the market.

The energy sellers will always want to maximize their profit. Considering the energy price associated with the prosumer offer $E_{offer}$ and the average energy price $avg_{price}(C_k)$ at the coalition level the prosumers preferences will have different values $v$:

$$v(C_k, E_{offer}) = \begin{cases} 1, & if\ avg_{price}(C_k) > price(E_{offer}) \\ 0, & if\ avg_{price}(C_k) > (price(E_{offer}) - \Delta_{price}) \\ -1, & if\ (price(E_{offer}) - \Delta_{price}) < avg_{price}(C_k) \end{cases} \quad (8)$$

where $\Delta_{price}$ is a price variation accepted by the seller in relation to its offer.

The buyers will always want to minimize their costs. Considering the energy price put on the bid $E_{bid}$ and the average energy price of the coalition their preferences will have different values:

$$v(C_k, E_{bid}) = \begin{cases} 1, & if\ avg_{price}(C_k) < price(E_{bid}) \\ 0, & if\ avg_{price}(C_k) < price(E_{bid}) + \Delta_{price} \\ -1, & if\ price(E_{bid}) + \Delta_{price} < avg_{price}(C_k) \end{cases} \quad (9)$$

The price preferences for joining the coalition are used to calculate the hedonic values of the coalition:

$$V_{price}(C_k) \begin{cases} \sum_{i=1}^{n} v(C_k, E_{offer,p_i}), \forall p_i \in C_k, p_i = seller \\ \sum_{i=1}^{n} v(C_k, E_{bid,p_i}), \forall p_i \in C_k, p_i = buyer \end{cases} \quad (10)$$

where $E_{bid,p_i}$ are energy bids of the buyers and $E_{offer,p_i}$ are the offers of the sellers belonging to the coalition $C_k$.

### 3.2. Coalitions creation and optimization

A genetic algorithm is used to determine the coalitions for P2P energy trading while considering the prosumer's social relationships and price preferences. It allows us to create coalitions of sellers and buyers based on social preferences during a market session and then to use the coalitions to match the demand and the offer. In the genetic algorithm, an individual is represented as a set of potential coalitions of prosumers either sellers or buyers:

$$I = \{C_1, \ldots, C_n\}\ where\ C_k = \{p_k | p_k = seller \oplus buyer\} \quad (11)$$

The prosumers in the coalitions of an individual are fetched from the market session:

$$C_1 \cup \ldots, C_n = all\ prosumers\ of\ the\ market\ session \quad (12)$$

An objective function is defined to assess the quality of an individual aiming to maximize the number of preferences satisfied among the players participating:

$$f(I) = (\sum_{i=1}^{n} w_i * |f_{pref}(C_i) - z_i^*|^m)^{1/m} \quad (13)$$

where $n = |I|$ is the number of coalitions composing an individual, $m$ is a control parameter that determines the type of distance used and $w_i$ is a weight vector for the relative importance of each preference function $f_{pref}$ of a coalition ($C_i$) of the individual and $z_i$ is the ideal value of the function. The weight vector has non-negative values that sum up to 1:

$$\sum_{i=1}^{n} w_i = 1 \quad (14)$$

The hedonic score for a coalition $C_i$ is computed as the sum of the hedonic values for all players $p_k$ in the coalition depending on its type (i.e., coalitions of sellers or coalitions of buyers):

$$f_{pref}(C_i) = V_{rel}(C_i) + V_{price}(C_i) \quad (15)$$

To generate the initial population, we will consider all prosumers' orders in the market session (i.e., either offer or bid for energy), and only those orders that have an energy quantity above a certain threshold $\Gamma$ will initiate coalitions:

$$E_{order,p_i} > \Gamma, E_{order,p_i} = \begin{cases} E_{offer,p_i}, p_i = seller \\ E_{bid,p_i}, p_i = buyer \end{cases} \quad (16)$$

The initial coalitions will be populated by distributing all the other orders equally to them. This step is performed multiple times, with a random distribution of orders, to explore different configurations of coalitions by keeping the initial set of orders consistent while varying the composition of the coalitions.

After generating the initial population of chromosomes (see Algorithm 1 in Figure 1), each chromosome is evaluated by computing the fitness function, using relation 13, while considering several factors such as the hedonic scores of the individual's coalitions, control parameters, and the weight vector. Then a selection process is employed to identify the most suitable parents for the new population based on their fitness scores. The process is repeated for a predetermined number of iterations, and during each iteration, it remembers the individual from all the resulting populations with the best global fitness value (lines 4-6). The process of updating the population uses a tournament selection procedure to select the two most fitted individuals from the population (lines 9-10). Then we will perform the reproduction process, where the selected parents are used to create a new generation of chromosomes using the crossover and mutation operators. The crossover operation is used to diversify the population (line 11) and is performed by swapping randomly between two individuals. The coalitions that correspond to this chosen point in both parent individuals will be selected and we will look for prosumers in those coalitions, with more enemy relationships than neutral / friendship.

---

**ALGORITHM 1: Prosumers hedonic coalitions**

**Inputs:** $P = \{p_1, p_2, \ldots, p_n\}$ the set of prosumers with energy bids or offers submitted in the market session, $\rho$ the maximum number of iterations, $\Gamma$ threshold for initiating coalitions

**Outputs:** $I_{best}$ — the best individual holding the optimal coalitions created

**Begin**
1. $N_c = COMPUTE\_COALITION\_NUMBER\ (P, bids, offers, \Gamma)$
2. $H_{Population} = GENERATE\_INITIAL\_POPULATION\ (P, Nc)$
3. $Fitness\ (H_{Population}) = \emptyset$
4. **Foreach** $I$ in $H_{Population}$ **do**
5.     $f_I = COMPUTE\_FITNESS\ (I, W_I, Z_I, m)$
6.     $Fitness\ (H_{Population}) = Fitness\ (H_{Population}) \cup f_I$
7. **End Foreach**
8. **For** i = 0, to $\rho$ **do**
9.     $parent_1 = TOURNAMENT\_SELECTION\ (H_{Population}, Fitness\ (H_{Population}))$
10.    $parent_2 = TOURNAMENT\_SELECTION\ (H_{Population}, Fitness\ (H_{Population}))$
11.    $Offsprings = CROSSOVER\ (parent_1, parent_2)$
12.    **Foreach** $offspring$ in $Offsprings$ **do**
13.      $offspring^* = MUTATE\ (offspring)$
14.      $f_{offspring^*} = COMPUTE\_FITNESS\ (offspring^*, W_{offspring^*}, Z_{offspring^*}, m)$
15.      $H_{Population} = UPDATE(H_{Population},\ offspring^*)$
16.    **End Foreach**
17.    $I_{best} = UPDATE\_BEST\_INDIVIDUAL\ (H_{Population}, Fitness\ (H_{Population}))$
18. **End for**
19. **return** $I_{best}$

**End**

*Figure 1. Create and optimize hedonic coalition using genetic algorithm.*

---

Swapping prosumers between coalitions might lead to cases in which an individual will not contain all energy orders but will have some duplicated orders. The fitness function for individual evaluation considers the diversity of an individual and will be greater for an individual with a higher diversity. The fitness function considers the relationship between the prosumers, so will favor the individuals with more friendship relations. The movement of prosumers can lead to situations where an individual might not have all the distinct energy orders, some orders are duplicated, or some are absent. The fitness function used for evaluating individuals considers their diversity. Individuals with a greater variety of prosumers and energy orders are assigned a better fitness score. Also, it favors individuals with more friendship relations among their members, the social relationships being important factors in determining an individual's fitness. However, the duplication of prosumers in a coalition will interfere with the fitness function computation because their relationships are also duplicated. Thus, we are not considering the duplicate relationships in the fitness function evaluation. In the case of the mutation operator (lines 12-16), we have selected the two most unstable coalitions in the individual. The instability is computed based on the relationships between prosumers in the coalition. The ones with the worst relationship scores will be considered. The goal is to guide the search space exploration towards coalitions with more favorable relationships. The mutation operator improves the overall quality of the solutions and avoids getting stuck in local optima. Our approach involves swapping prosumers between coalitions with significant differences in hedonic score and ideal score. Only prosumers in an enemy relationship are selected for exchange. The fitness value of offspring is then calculated by removing duplicates.

Finally, the population of individuals is updated with the newly generated offspring. To update the population (see Algorithm 2 in Figure 2), we use a strategy that considers not only the fitness values but also the contribution to population diversity.

---

**Algorithm 2: Update population of hedonic coalitions**

**Inputs:** *offspring*, $H_{Population}$ – hedonic population at the start of the current iteration
**Outputs:** $H_{Population}$ - updated hedonic population at the end of the current iteration
**Begin**
1. **Foreach** $I$ in $H_{Population}$ **do**
2.    **if** $(f_{offspring} > f_I)$ $H^*_{Population} = H_{Population} \cup \{I\}$
3. **End foreach**
4. $CD = \emptyset$
5. **Foreach** $I^*$ in $H^*_{Population}$
6.    $CD^* = \textbf{COMPUTE\_DIVERSITY}$ $(I^*, H^*_{Population})$;
7.    $CD = CD \cup CD^*$
8. **End foreach**
9. $c_{min} = SELECT(H^*_{Population}, MIN(CD))$
10. $H'_{Population} = H^*_{Population} - \{c_{min}\}$
11. $CD' = \textbf{COMPUTE\_DIVERSITY}(offspring, H'_{Population})$
12. **if** $(CD' > c_{min}) \& (\neg Duplicate(offspring, H^*_{Population})$
13.    $H^*_{Population} = \textbf{REPLACE}$ $(H^*_{Population}, c_{min}, offspring)$
14. **else**
15.    **Foreach** $I$ in $H^*_{Population}$ **do**
16.      $Fitness(H^*_{Population}) = Fitness(H^*_{Population}) \cup f_I$
17.    **End Foreach**
18.    $I_{worst} = \textbf{SELECT\_WORST}(H^*_{Population}, Fitness(H^*_{Population}))$
19.    **if** $(\neg Duplicate(offspring, H^*_{Population})$
20.      $H^*_{Population} = \textbf{REPLACE}(H^*_{Population}, I_{worst}, offspring)$
21.    **endif**
22. **endif**
23. **return** *Population*
**End**

*Figure 2. The hedonic population updating process.*

This approach is applied to each offspring individually and aims to replace an individual in the population with a lower fitness value and less diversity contribution with the offspring. The strategy begins by identifying the chromosome with the least diversity contribution, among those with lower fitness values than the offspring. The diversity contribution of a chromosome is calculated using the similarity between the chromosome and its closest neighbor in the population:

$$cd(I, H_{Population}) = \min_{I' \in Population, I' \neq I} d(I, I') \quad (17)$$

where $d(I, I')$ is the distance between the two individuals $I$ and $I'$ computed as:

$$d(I, I') = \frac{\sum_{i=1}^{n} d(C_i^I, C_i^{I'})}{n} \quad (18)$$

In formula (18) refers to the number of coalitions that make up a chromosome (in our case it is assumed that all chromosomes have the same dimension), and $d(C_i^I, C_i^{I'})$ is the distance between corresponding pairs of coalitions from the two individuals computed using the Jaccard Index:

$$JacccardIndex(C_i^I, C_i^{I'}) = \frac{C_i^I \cap C_i^{I'}}{|C_i^I| + |C_i^{I'}| - |C_i^I - C_i^{I'}|} \quad (19)$$

where $C_i^I, C_i^{I'}$ are two coalitions of prosumers, $C_i^I \in I \text{ and } C_i^{I'} \in I'$.

Within the current population, the individuals having the lowest fitness and diversity contribution scores are identified. When an offspring displays both better fitness and increased genetic diversity compared to the least fit individual in the current population, that offspring is introduced to the population. In cases where the offspring doesn't substantially enhance diversity but still outperforms the least fit individual in terms of fitness, the offspring takes the place of the least fit individual. This ensures an ongoing improvement in the population's overall fitness, even if genetic variety isn't significantly boosted.

After updating the population, we repeat all the steps above for a predefined number of iterations, keeping track of the individual with the best global fitness value in all iterations. For each coalition belonging to the best individual, the total amount of energy is determined by summing the amount provided by the prosumers in their bids and offers while the price is determined as an average of the price in the bids and offers:

$$E_{order}(C_k) = \begin{cases} (\sum_{i=1}^{n} E_{offer,p_i}, avg(price_{p_i})), p_i \in C_k, p_i = seller \\ (\sum_{i=1}^{n} E_{bid,p_i}, avg(price_{p_i})), p_i \in C_k, p_i = buyer \end{cases} \quad (20)$$

## 4. Evaluation Results

To evaluate the hedonic game solution, we have used a P2P energy flexibility market that was introduced in [9]. The flexibility market operates over a blockchain network (see Figure 3) which keeps in a tamper-proof manner the energy transactions of prosumers, the bids and offers submitted in a market session, etc. The market is set up at an energy community level and each prosumer will have his smart contract deployed on the private blockchain network to manage their interactions. Using monitored energy data of prosumers, their flexibility is assessed and digitized using ERC721 and ERC20 tokens [42]. The market is operated using specific smart contracts to handle market registration and verifications, keeping track of session type, the energy flexibility orders that are placed for each session, as well as resulting transactions from these orders. We modeled the transfer of energy flexibility over the P2P energy market using monitored energy data from 14 prosumers, 4 energy buyers, and 10 energy sellers [9]. The energy profiles of prosumers are presented in Table 1.

*Table 1. Energy profiles of prosumers used for order generation on the P2P market.*

| Prosumers | | | |
|---|---|---|---|
| Buyers | | Sellers | |
| Ids | Energy Interval [Min-Max] | Ids | Energy Interval [Min-Max] |
| 1 | [1, 14] | 4 | [0, 9] |
| 2 | [1, 4] | 5, 6 | [0, 11] |
| 3 | [1, 20] | 7 | [0, 12] |
| 4 | [2, 15] | 8 | [0, 14] |
| | | 9 | [0, 16] |
| | | 10 | [0, 17] |
| | | 11 | [1, 23] |
| | | 12 | [1, 19] |
| | | 13 | [1, 20] |

The following actions are carried out for P2P energy trading. The prosumers register with the market manager smart contract and using their contracts will place day-ahead offers or bids in the market session (i.e., for each hour of the next day). After all the orders are placed, the P2P matching in transactions is done with our hedonic game solution. We have implemented it as a layer 2 solution to address the cost issues for running complex algorithms on-chain. The generated transactions are stored in the blockchain, and during the next day the commitment energy and financial settlement is carried out. The prosumers delivered energy values based on the monitored values from power meters are compared to the committed ones in energy transactions and tokens are delivered to their wallets.

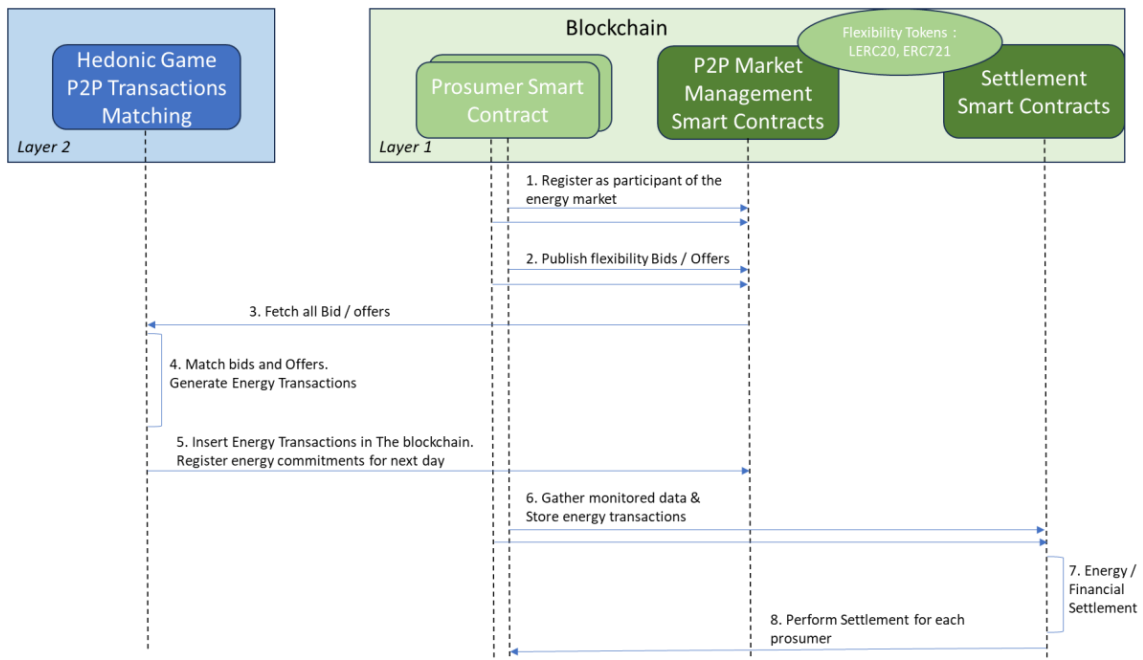

*Figure 3. P2P Energy Market Operation and Hedonic Game Solution Integration*

Initially, we assess the viability of the heading game solution in creating P2P energy transactions while adhering to social limitations. The parameters under examination include the reduction in the desired number of transactions and the increase in the amount of energy transferred. They are important to reduce the overheads while increasing the system's throughput. Moreover, we are interested in the prices at which prosumers will trade energy and how much profit they can make. The energy bids and offers are used as input data for our Hedonic Game solution and assess the amount of energy transferred and if the balance between the demand and generation is ensured. The results are reported in Figure 4 for three hours selected for the next day. The hedonic game solution for P2P trading achieves an equilibrium between community-level demand and generation, despite the variations in the energy transferred during each hour.

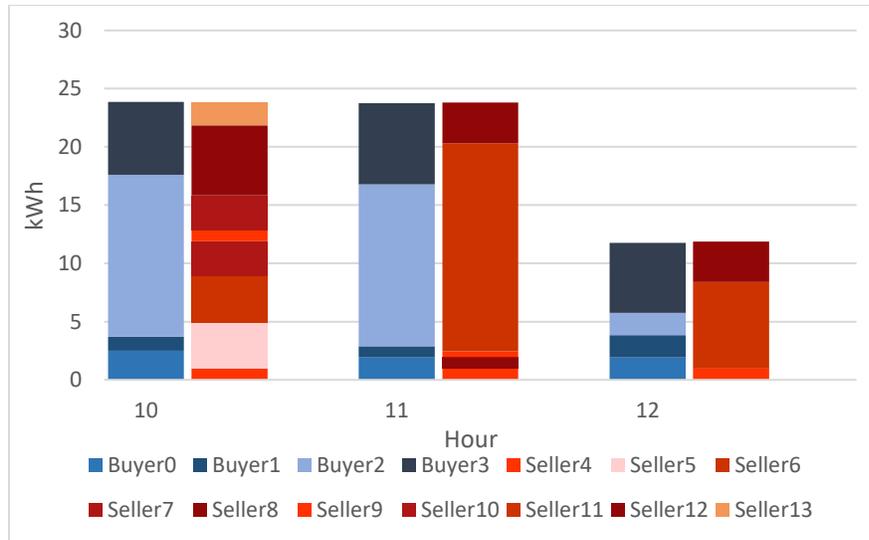

*Figure 4. Energy transactions for hours 10,11,12*

Moreover, the solution generates a relatively low number of energy transactions for the market session, proportional to the total amount of energy and the number of orders that were placed for each hour.

Figure 5 shows the prices for trading energy in the selected hours concerning the prosumer ID of each participant in the market session, its role either as buyer or seller, and the price in Gwei. The transactional prices are scattered on the graph, as they depend on the coalitions in which buyers and sellers are matched by the hedonic game solution proposed. Moreover, the prices at which prosumers trade energy have a similar average value for the selected hours.

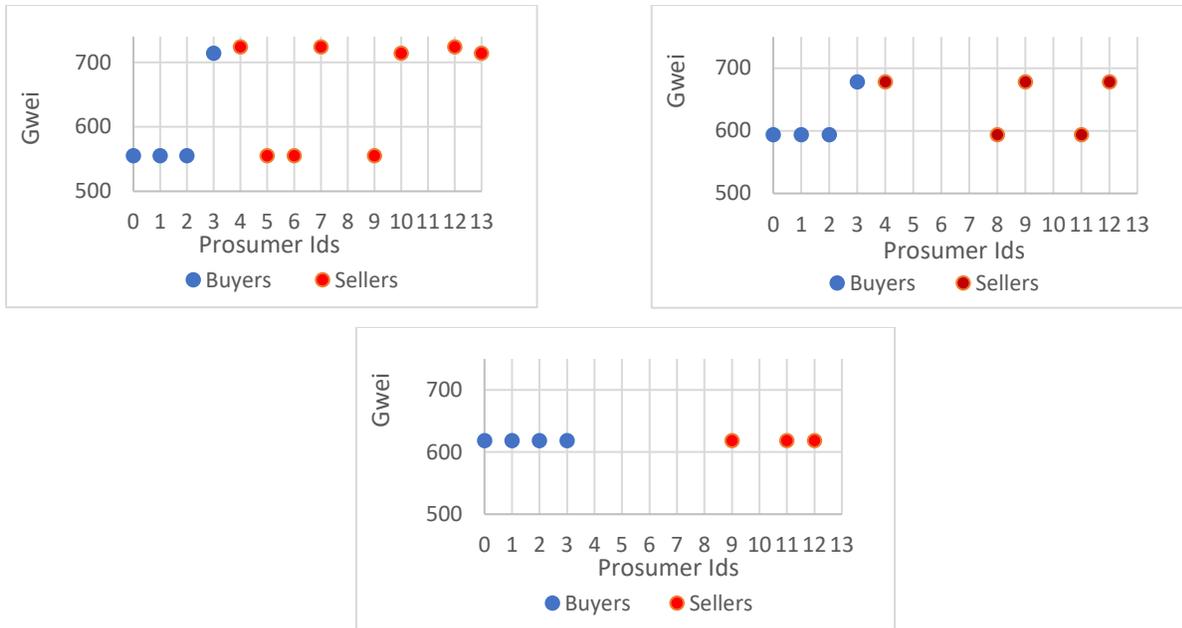

*Figure 5. Energy Transaction Prices for hours 10, 11 and 12*

Next, we will analyze the coalitions from the perspective of the social factors considered. We assess the number of friendship-neutral and enemy relations from the coalitions created using the hedonic game (see Table 2). If we take a closer look at each coalition, namely at the prosumers forming them, we can see a tendency to have good relationships. Most of the prosumers in each coalition have friendly relationships, with small exceptions due to the small sample size of the scenario.

*Table 2. Social relationships among prosumers in coalitions.*

| Hour | Prosumers in Coalition | Friendship | Neutral | Enemy |
|---|---|---|---|---|
| 10 | <6,13> | 0 | 1 | 0 |
| | <7,12,4> | 3 | 0 | 0 |
| | <1,2> | 2 | 0 | 0 |
| 11 | <6,13,5> | 1 | 2 | 0 |
| | <4,12,7> | 3 | 0 | 0 |
| | <11,9> | 1 | 0 | 0 |
| | <3,1,2> | 3 | 0 | 0 |
| 12 | <5,6,7> | 2 | 0 | 1 |
| | <11,13,9> | 2 | 1 | 0 |
| | <4,12,8> | 2 | 1 | 0 |
| | <2,1,3> | 3 | 0 | 0 |

For a comprehensive understanding of the entire market dynamics and coalition formation throughout a market session, we've used the hedonic score defined in relation (7) (see Figure 6). This index shows the collective social status of all coalitions during a given hour. Positive values above 0 indicate favorable social interactions among prosumers within the coalitions, with higher values denoting stronger social bonds.

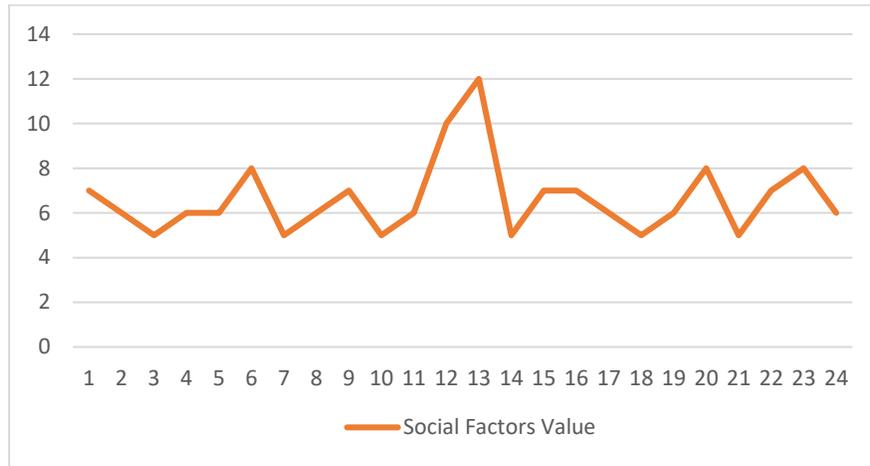

Figure 6. Social factor value for the entire market session coalitions of prosumers

We have evaluated the proposed solution in a more complex scenario involving a higher number of prosumers and compared the results with other game theory-based solutions (G-T) [17]. For this goal, we have generated the energy profiles for 48 prosumers (i.e., 24 energy buyers and 24 energy sellers) and a random energy price for each prosumer. The first aspect compared was the number of energy transactions generated under increased prosumer participation during the market session. We are using a randomized value for the parameters, and the results in terms of the transaction amounts that were completed in an hour of the next market session are presented in Figure 7.

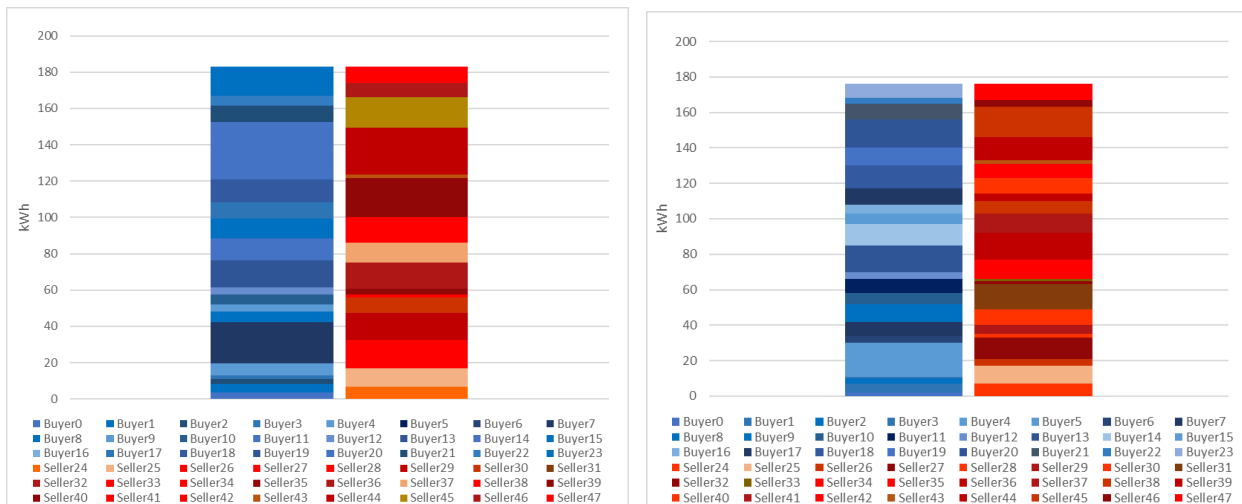

Figure 7. Energy transactions among coalitions: (left) hedonic game and (right) G-T.

The results show that the Hedonic game and G-T are capable of matching supply and demand in P2P energy trading. However, the Hedonic game approach commits more energy to transactions, potentially increasing market liquidity. Both solutions have similar prices, but G-T outperforms the Hedonic game solution due to lack of social factors consideration (see Figure 8, the energy price is reported in Gwei).

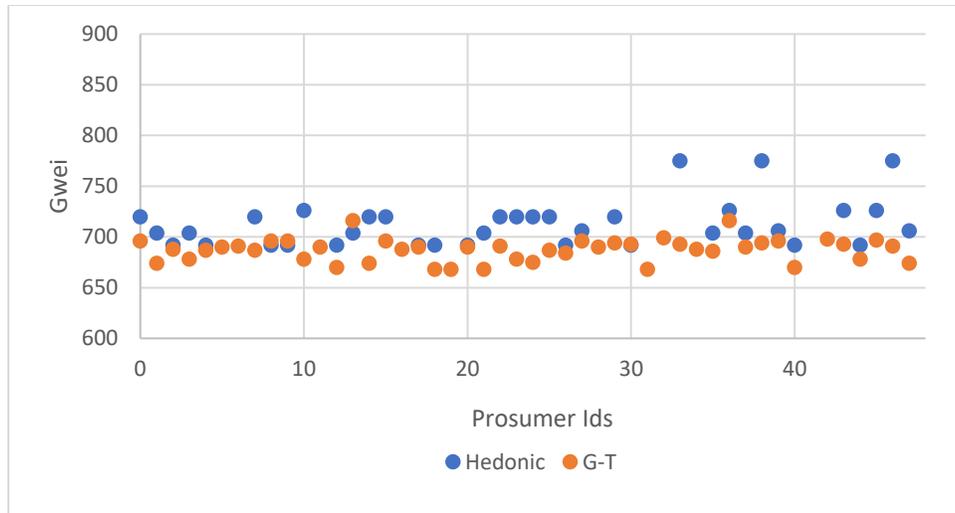

*Figure 8. Energy price comparison between solutions*

## 5. Discussion

In this section, we discuss the impact of several relevant parameters of the hedonic game solution outcomes. The parameters analyzed are the number of coalitions, the distribution of social relationships in the community, and the fitness function value. For all these parameters, we have used the same prosumers energy profiles and price values, and we varied each parameter three times and compared the results obtained.

First, we analyzed the number of coalitions' impact on the total amount of energy transferred in an energy market session, as well as on the energy prices. We have varied the threshold set (Γ see relation 16) for initial coalitions formation and for joining existing coalitions allowing the creation of (i) 4 coalitions – 2 buying and 2 selling energy, (ii) 13 coalitions – 6 buying and 7 selling energy, and (iii) 23 coalitions – 11 buying and 12 selling.

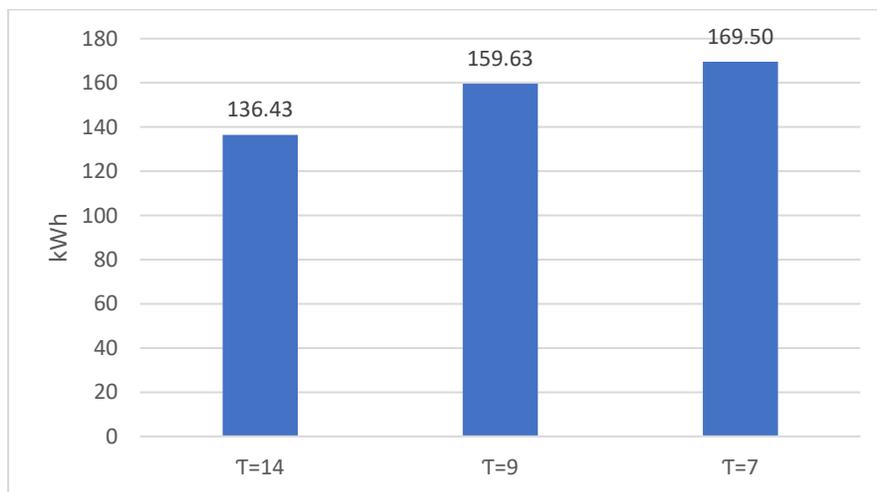

*Figure 9. The total amount of energy traded for different coalition formation thresholds.*

Figure 9 shows that having a lower Γ value in the hedonic game solution leads to more coalitions and more energy being transacted in an energy market session. With fewer coalitions, there is a significant reduction in the amount of transferred energy, while above a Γ value, the amount of energy transferred only slightly increases. This is because the hedonic game solution that receives coalitions after the best individual is generated performs a more accurate matching if there are more orders to be matched. Increasing the total number of coalitions from 4 to 13 resulted in a 15% increase in transferred energy while increasing it to 23 only increased transferred energy by 5%. In terms of energy price variation

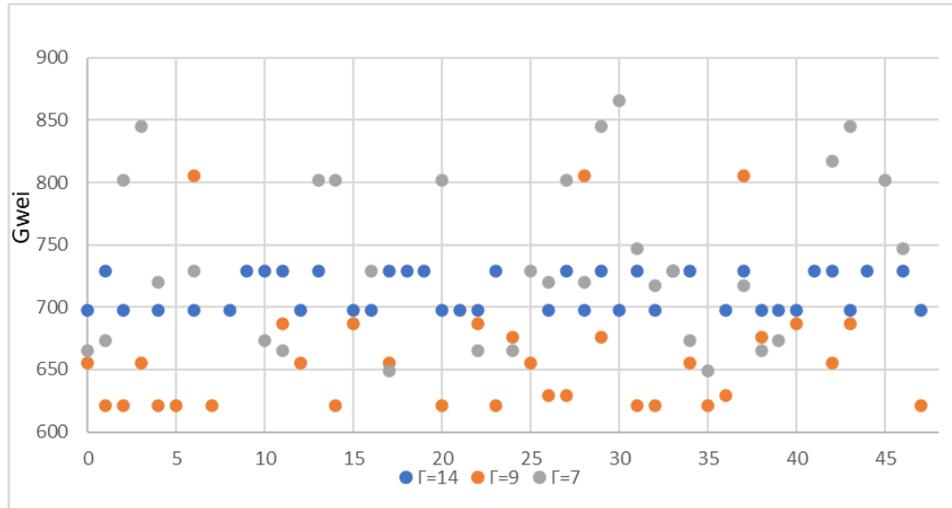

Figure 10. Price analysis when varying the threshold for creating coalitions.

Figure 10 shows that the more coalitions created (i.e., lower threshold), the more scattered the prices will be because, after choosing the best individual, each coalition forming this individual will become a new order and it will be matched by the hedonic game solution.

The second parameter analyzed is the social relationship distribution in the community. More precisely, we have varied the proportions of enemy, neutral, and friendship relationships between prosumers in the energy community and determined the impact on the amount of energy traded in a market session (see Figure 11).

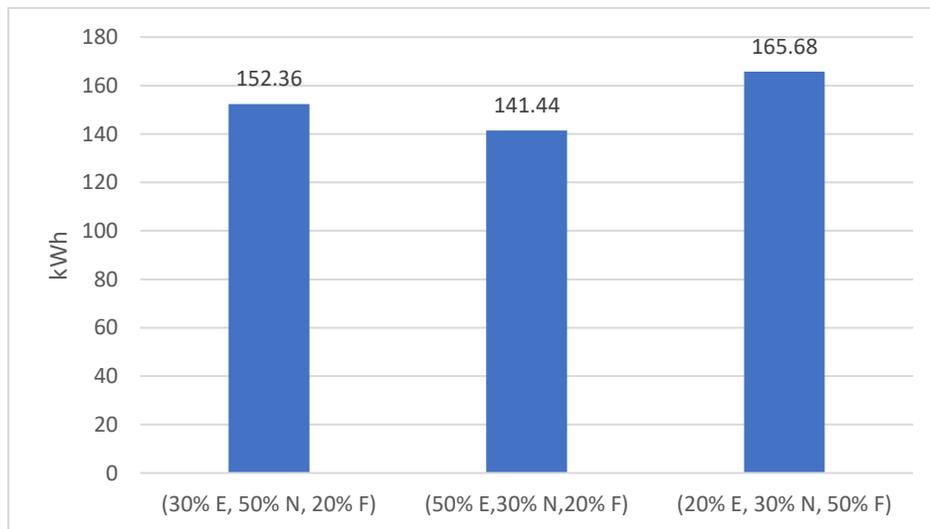

Figure 11. Total amount transacted when varying the relationships between prosumers.

The amount of energy transacted is influenced by the fitness function that considers the relationships between prosumers. The selection of the best individual is affected if there is an increased number of neutral relationships as it cannot differentiate between social relations among prosumers. However, a community with more friendship relationships has an increase in total energy transacted by about 10%. The clear definition of social relationships among prosumers is important as largely neutral prosumers can eventually lead to decreased energy trading. The fitness function is influenced by the weight of coalitions based on the relationship status of the prosumers forming it, and the energy price variation in the submitted bids and offers. We have used the same distribution of social relationships among prosumers and varied the weights for the coalitions in the individuals. All coalitions that have a positive total relationship value of prosumers forming it, will have their weights add up to 0.6. The coalitions that have a neutral value of relationships (equal number of friendship and enemy relationships), will have a sum of weights of 0.2. Finally, coalitions within the individual that have a negative sum of relationships within it, have the sum of weights equal to 0.2. Thus, the fitness function will promote the coalitions that have prosumers who are mostly friends, so these individuals will be more favored and will have a greater contribution towards finding the most fitted individual.

Figure 12 shows the total energy transacted in the market session. In instances where predominantly unfavorable interactions exist among prosumers, the energy transfer is reduced by 25% in comparison to the other scenarios. The energy prices, even though are scattered for all cases, are better with 4% for coalitions with predominant friendship social relationships compared with the ones with predominant neutral ones (see Figure 13).

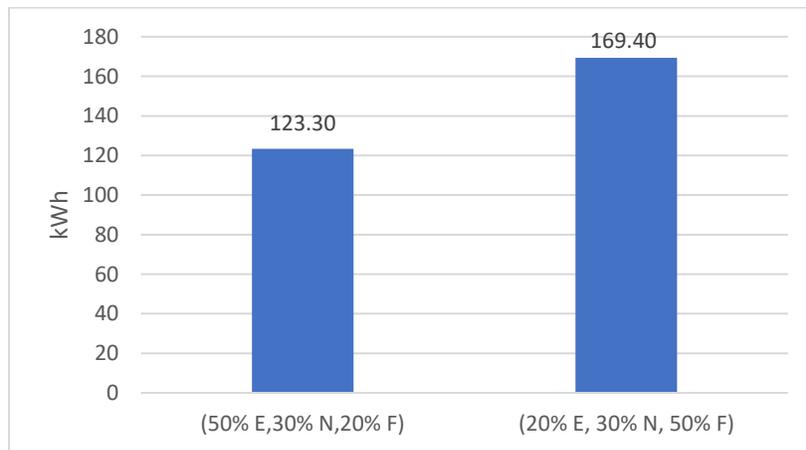

Figure 12. Total transacted energy amount when promoting the coalitions with good social relationships.

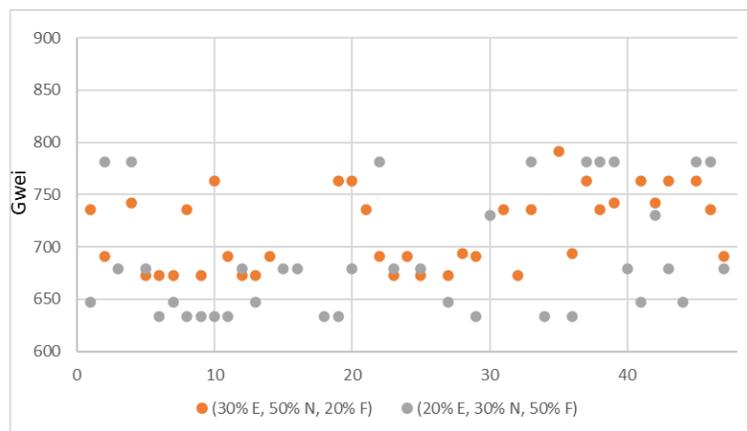

Figure 13. Energy price for mostly neutral and positive social relations.

# 6. Conclusions

In this paper, we have presented a prosumer hedonic game-based coordination and cooperation model for P2P energy flexibility trading in an energy community of prosumers. The hedonic game model proposed addresses the creation of coalitions of prosumers, considering their preferences expressed as social connections with other peers, and uses genetic heuristics to optimize the prosumers' coalitions to balance the energy demand and supply within the community. The implementation and evaluation were done with a P2P energy flexibility market based on blockchain and smart contracts.

The evaluation results show the effectiveness of our solution that integrates hedonic game, genetic heuristic, and blockchain P2P energy market in considering both social relationships and price preferences of prosumers when forming coalitions, ultimately contributing to more balanced energy demand and supply within the community. Also, the results in the context of an energy community are promising, highlighting the potential of the hedonic game model in enhancing the effectiveness of P2P energy trading by increasing the participation of prosumers and the total amount of energy transacted in a market session. Finally, it shows the importance of the social dimensions of P2P energy transactions, providing insights into the role of cooperation and social dynamics in community-driven energy trading systems.

For future work, we plan to consider more in-depth the role of prosumers' behavior in their actions in a decentralized energy market. The prosumers may have conflicting objectives and strategies, not just to maximize the community's benefits, and we plan to investigate further and see how these behaviors might influence P2P energy trading.

# Acknowledgement

This work has been conducted within the DEDALUS project grant number 101103998 funded by the European Commission as part of the Horizon Europe Framework Programme.